\documentclass[twocolumn,showpacs,preprintnumbers,amsmath,amssymb]{revtex4}
\usepackage{graphicx}
\begin{document}

\date{\today}
\vspace{2.7in}

\title{Can we move photons?}

\author{Oleg L. Berman$^{1}$, Roman Ya. Kezerashvili$^{1,2}$, and Yurii E.
Lozovik$^{3,4}$}
\affiliation{\mbox{$^{1}$Physics Department, New
York City College of Technology, The City University of New York,}
Brooklyn, NY 11201, USA \\
\mbox{$^{2}$The Graduate School and University Center, The
City University of New York,}
New York, NY 10016, USA \\
\mbox{$^{3}$Institute of Spectroscopy, Russian Academy of Sciences,
142190 Troitsk, Russia} \\
\mbox{$^{4}$Moscow Institute of Physics and Technology (State
University), 141700, Dolgoprudny, Russia }\\}

\begin{abstract}

The drag effects in the system of spatially separated electrons and
excitons in coupled quantum wells (CQWs) embedded in an optical
microcavity are predicted. It is shown that at low temperature an
electron current induces the polariton flow, therefore, a transport
of photons along the cavity. However, the superfluid polariton
component does not contribute to the electron drag.  The
polariton-electron at the low temperatures and exciton-electron at
the high temperatures drag coefficients are presented. It is shown
that the drag coefficients increase when temperature increases. We
discuss possible experiments for the observation of the
electron-polariton drag effect.

\vspace{0.1cm} Key words: Drag effect; Exciton-polaritons;
Superfluidity
\end{abstract}


 \maketitle

Recently   properties of 2D excitonic polaritons~\cite{Agranovich}
in a quantum well (QW) embedded in an optical microcavity have been
the subject of theoretical and experimental studies (see
Refs.~[\onlinecite{book,Snoke_text}] and references therein). The
system under study consists of two Bragg mirrors, forming an optical
microcavity, and quantum well embedded within the cavity at the
antinodes of the confined optical mode. The resonant exciton-photon
interaction results in the Rabi splitting in the excitation spectrum
and formation of upper and lower polaritons, which are the
superpositions of the excitons and cavity photons. It was assumed
that the modes of the microcavity photon and exciton in a QW were in
resonance at $k_{||} = 0$, where $k_{||}$ is the component of the
wave vector along the microcavity.
 The extremely  small effective mass of the lower polariton  results in a
 high critical  temperature for superfluidity~\cite{science,Lit}. Remarkable success
was achieved in the observation of condensation, coherent effects
and anomalous transport properties  in the microcavity polariton
system~\cite{Kasprzak,Balili_prb,Amo1,Amo2}. The unambiguous, direct
observation of the superfluidity of the cavity polariton system is
still an actual and essential task.

In the present Letter we predict the drag effects of the cavity
polaritons and electrons, and  analyze the manifestation of the
polariton superfluidity in these phenomena. We consider two
neighboring quantum wells  embedded in an optical microcavity: the
first QW is occupied by 2D electron gas (2DEG) and the second QW is
occupied by the excitons created by the laser pumping.

At low temperature  $k_{B}T \ll \hbar\Omega_{R}$, where $\hbar
\Omega_{R}$ is Rabi splitting, and $k_{B}$ is Boltzmann constant,
and at the resonance of excitons with cavity photons, the exciton
polaritons are formed by the quantum superposition of the excitons
with the cavity photons~\cite{text2}. By focusing laser pumping in
some region of the cavity the gradient of excitons and exciton
polaritons can be generated. These gradients induce the polariton
and exciton currents, and in a turn the normal (non-superfluid)
component of moving exciton liquid  drags the electrons in the
neighboring QW due to the electron-exciton interaction. So the
electric current would be generated by the current of the normal
component of exciton polariton liquid.

In another scenario by applying electric voltage in the QW with 2DEG
the electronic current is induced, and this current drags the normal
(non-superfluid) component of exciton liquid in the  neighboring QW.
There is the quantum superposition of the excitons and the cavity
photons. So the cavity photons can be dragged and, thus, controlled
by the electric voltage.
 Besides, possible applications of the control of photons or electrons by the polariton-electron drag can be used for study of the properties and phases
 in the polariton and exciton system, particularly, the superfluidity of the system (in some analogy with the Coulomb drag effects
 in the electron-hole system, see Refs.~[\onlinecite{Lozovik,Pogrebinskii,Price,Gramila,Vignale}] and references
 therein). Interestingly enough to mention that the first time drag effects were studied in the electron-phonon system in 1945 in Ref.~[\onlinecite{Gurevich}].

 Note that at high temperature, $k_{B}T \gtrsim \hbar\Omega_{R}$, the polariton states are occupied mainly far from the polariton resonance, and in these
 states the exciton-photon quantum superposition is negligible. Thus, at high temperatures only the exciton-electron drag is
 essential, and the exciton current can induce the electron current as well as the electron current can produce  the exciton current.

We introduce the drag coefficients $\lambda _{p}$ and $\lambda
_{ex}$ for electrons in the 2DEG dragged by the moving polaritons
and exitons, respectively. For the case when the electric field is
applied to the system of electrons in the QW we introduce the drag
coefficients $\gamma _{p}$ and $\gamma _{ex},$ respectively, for
polaritons and excitons dragged by the electron current. In the
two-layer system there are a current of electrons and
current of polaritons or excitons. The current of polaritons or excitons is $%
\mathbf{i}_{i}=n_{i}\mathbf{v}_{i}$, where $n_{i}$ and $v_{i}$ are
density and average velocity, and the index $i$ is defined as $i =
ex$ for excitons and $i = p$ for polaritons. The electron current $\mathbf{j}%
=-en_{el}\mathbf{v}_{el}$, where $n_{el}$ is\textbf{\ }the density
 and $\mathbf{v}_{el}$ is the average velocity of electrons in the electron layer.  These currents can be expressed in
terms of the density gradient $\mathbf{\nabla }n_{i}$ in the
polariton or exciton subsystem, drag coefficients $\lambda _{i}$ and
$\gamma _{i},$ and external electric field $\mathbf{E}$ applied to
the 2DEG by the following matrix expression:
\begin{equation}
\left(
\begin{array}{c}
\mathbf{i}_{i} \\
\mathbf{j}%
\end{array}%
\right) =\left(
\begin{array}{cc}
-D_{i} & \gamma _{i} \\
\lambda _{i} & en_{el}D_{e}%
\end{array}%
\right) \cdot \left(
\begin{array}{c}
\mathbf{\nabla }n_{i} \\
\mathbf{E}%
\end{array}%
\right) \ , \label{matrix}
\end{equation}%
where $D_{i}$ is the polariton or exciton diffusion coefficient and
$D_{e}$ is the mobility coefficient of the electrons. Only normal
component in the polariton subsystem is dragged by the electron
current, while the superfluid component is not dragged. Thus, the
appearance of the polariton superfluidity can be detected by the
electron-polariton drag effect.

Below in our calculations we assume that the exciton system is
dilute, satisfying to the following condition $n_{ex}a_{0}^{2}\ll
1$, where $n_{ex}$ is the quasistationary concentration of excitons,
created by the pumping, $a_{0}$ is the 2D exciton Bohr radius. The
condition $n_{ex}a_{0}^{2}\ll 1$ holds for the excitons at the
exciton densities up to $n_{ex}\approx 10^{12}\ \mathrm{cm}^{-2}$,
since in
GaAs/GaAsAl quantum well the exciton Bohr radius is in the order of $%
a_{0}\sim 10-50\ \mathrm{{\mathring{A}}}$. It is obvious to conclude
that the system of polaritons is also dilute, since the system of
excitons, that forms the polaritons, is dilute.

We obtain the drag coefficients $\gamma_{i}$ from the expansion of the polariton or exciton
current $\mathbf{i}_{i}$ by keeping the leading term linear by the electric field $%
\mathbf{E}$. This current is given by
\begin{equation}
\mathbf{i}_{i}=\frac{1}{M_{i}}\int \mathbf{p}n_{i}(\mathbf{p})\frac{%
s d^{2}p}{(2\pi \hbar )^{2}} \ ,  \label{cur}
\end{equation}
where $\mathbf{p}$ is the polariton or exciton momentum, $M_{i}$ is the effective mass of polariton or mass of exciton,
 $s = 4$ is the spin degeneracy, and $n(\mathbf{p})$ is the distribution function of the
excitons or quasiparticle excitations in the polariton subsystem
\begin{eqnarray}
n_{i}(\mathbf{p})=n_{0i}(\mathbf{p})\left[1 +    (1+n_{0i}(%
\mathbf{p}))g(\mathbf{p}) \right]\ .  \label{distex}
\end{eqnarray}
 In (\ref{distex}) $g(\mathbf{p})$ is the
contribution to the quasiparticle distribution function
corresponding to the nonequilibrium correction due to the gradient
of the chemical potential of polaritons (or excitons at high $T$)
which was found by solving the kinetic equations. In (\ref{distex})
$n_{0i}(\mathbf{p})$ is the Bose-Einstein distribution function of
the quasiparticles in the polariton or exciton subsystems at the
equilibrium:
\begin{eqnarray}
\label{bed}
n_{0i}(\mathbf{p})=\left( \exp ([\varepsilon_{i}(\mathbf{p})-\mu _{i}]/(k_{B}T)])-1\right) ^{-1} \ .
\end{eqnarray}
In~(\ref{bed}) the energy spectrum of the quasiparticles and
chemical potential are given within  the Bogoliubov approximation.
However, for non-interacting excitons in~(\ref{bed}) the single
particle spectrum is $\varepsilon_{ex} = p^{2}/(2M_{ex})$.
  The effective mass of a polariton is
\begin{equation}
M_{p} = 2\left( M_{ex}^{-1}+\frac{cL_{C}}{\sqrt{\epsilon }\hbar \pi
}\right) ^{-1}\ ,  \label{Meff}
\end{equation}
where $M_{ex} = m_{e}+m_{h}$ is the exciton mass, $L_{C}$ is the
length of optical microcavity, and $\epsilon $ is the dielectric
constant (see, e.g, Ref.~[\onlinecite{Berman_L_S}] and references
therein).

In a many-particle system of excitons and electrons we replaced the
bare pair exciton-electron interaction by the effective interaction
$W_{eff}$~\cite{Kulakovskii} corresponding to the exciton-electron
interaction screened by the electron-electron interactions in 2DEG
in the random phase approximation.

We have obtained the polariton distribution function~(\ref{distex})
by solving the kinetic equations and using the exciton-electron
interaction vertex in the Born approximation. By substituting it in
(\ref{cur}), we find the following estimation for the drag
coefficient $\gamma _{p}$ at low temperatures $k_{B}T\ll \mu $ and
$k_{B}T\ll \hbar \Omega _{R}$ (we skip over the complicated
derivations and present only the simple estimation for $\gamma
_{p}$):
\begin{eqnarray}
\label{alpha_2} \gamma _{p}=\frac{1}{144}\left( \frac{21}{32}\right)
^{2}\frac{\pi
^{2}s^{2}e^{5}a_{0}^{6}\bar{\tau}_{n}\bar{\tau}_{e}}{\hbar
^{3}a_{e}^{4}\epsilon ^{2}D^{4}}\exp \left[ -\mu_{p} /(k_{B}T)\right] .
\end{eqnarray}
Note that the polariton relaxation time can be expressed in terms of
the  relaxation time of the normal (non-superfluid) component of
exciton liquid. So in (\ref{alpha_2}) we approximated  the
relaxation time of the normal component of exciton liquid and
electron relaxation time by their average values $\tau _{n}$   and
$\tau _{e}$, respectively. These average values
can be obtained from the exciton mobility $\tilde{\mu}_{ex}=e\bar{\tau}%
_{n}/M $ and the electron mobility $\tilde{\mu}_{e}=e\bar{\tau}%
_{2}/m_{e}$ presented in Fig.~1 and Fig.~2 in
Ref.~[\onlinecite{Walukiewicz}]. Since derivation $\gamma_{p}$ resulting in~(\ref{alpha_2}) implies
the condition $k_{B}T \ll \mu$ and $k_{B}T \ll \hbar \Omega_{R}$, we
can apply~(\ref{alpha_2}) for the temperatures below $\sim 20 \
\mathrm{K}$.

Let us mention that Eq.~(\ref{alpha_2}) was obtained by using the
regular Bogoliubov approximation for the weakly-interacting Bose gas
with no dissipation~\cite{text1}. We assume that the leakage of the
photons from the microcavity  is very small, and the system can be
 considered in the quasi-equilibrium.

For high temperature, $k_{B}T\gtrsim \hbar \Omega _{R}$, the
majority of polaritons occupy the upper polariton branch, where the
upper polariton mass is very close to the mass of exciton $M_{ex}$.
So at high temperature the polaritons are replaced by the free Bose
gas of excitons~\cite{Lit}.

We apply for the estimation of the exciton-electron drag coefficient
for the dilute exciton system the same approach as for the
polariton-electron drag coefficient for the dilute polariton system
 and present only the simple estimation:
\begin{eqnarray}  \label{alpha_3}
\gamma_{ex} &=& \frac{1}{144} \left(\frac{21}{32} \right)^{2} \frac{%
\pi^{2}s^{2}e^{5}a_{0}^{6}\bar{\tau}_{n}\bar{\tau}_{e}} {\hbar^{3}a_{e}^{4}%
\epsilon^{2}D^{4}}  \nonumber \\
& \times & \exp\left[ (k_{B}T - \mu_{ex})/(k_{B}T) \right] \ .
\end{eqnarray}

It is easy to see that at low temperatures in the very dilute
systems of the polaritons the collective excitations described by
the linear region in the spectrum give no contribution to
$\lambda_{p}$. In the quasiparticle approach  the normal component
of the polariton system can be calculated by the integration of the
Bose-Einstein distribution function~(\ref{bed}) for the
quasiparticles. Since $\lambda_{p}$ is directly proportional to
$\partial \mu _{i}/\partial n_{i}$~\cite{Nikitkov}, we obtain for
the polariton subsystem
\begin{eqnarray}
\label{der1}
\left( \frac{\partial n_{p}}{\partial \mu _{p}}\right) _{\mu _{p}=0}=%
\frac{sk_{B}T}{2\pi \hbar ^{2}c_{s}^{2}}\int_{0}^{\infty }\frac{e^{x}xdx}{%
(e^{x}-1)^{2}} \ ,
\end{eqnarray}
where $c_{s}$ is the sound velocity for the quasiparticles in the polariton subsystem~\cite{Berman_L_S}, and
\begin{eqnarray}
\label{npm3} \left. \left( \frac{\partial n_{ex}}{\partial \mu
_{ex}}\right) \right\vert
_{\mu _{ex}=\mu _{ex}^{(0)}}=\frac{sM_{ex}}{2\pi \hbar ^{2}}\left[ 1-\exp \left[ -%
\frac{2\pi \hbar ^{2}n}{sM_{ex}k_{B}T}\right] \right] \
\end{eqnarray}
for the ideal 2D Bose gas of excitons. The integral in the r.h.s.
of~(\ref{der1}) diverges. Therefore, we have $\partial
n_{p}/\partial \mu _{p}\rightarrow \infty $, which results in
$\lambda _{p}=0$ according to the fact that  $\lambda_{p}$ is
directly proportional to  $\partial \mu _{i}/\partial
n_{i}$~\cite{Nikitkov}. This result comes from the assumption that
for the very dilute Bose gas of polaritons we took into account only
the sound region of the collective excitation spectrum at small
momenta, and neglect almost not occupied regions with the quadratic
spectrum at large momenta and crossover dependence of the spectrum
at the intermediate momenta. Our approximation results in
 suppressed $\lambda_{p}$. Therefore, in the presence of
the superfluidity of polaritons, the polaritons moving due to their
density gradient almost do not drag electrons and there is
suppressed electron current induced by the polaritons. Hence, the
suppression of the dragged electric current in the electron QW can
indicate the superfluidity of the polaritons.

\begin{figure}[t] 
   \centering
  \includegraphics[width = 2.5in]{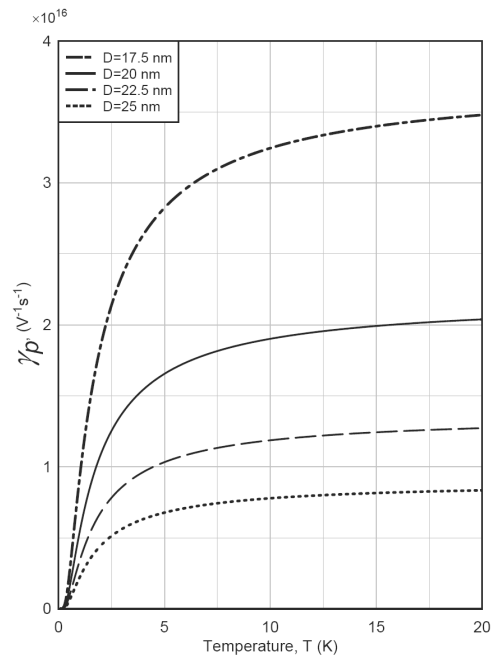}
\caption{ The drag coefficient $\gamma_{p}$ for the system of
spatially separated superfluid microcavity polaritons and electrons
 as a function of temperature $T$ for the different interwell separations $D$. }
   \label{gam_lt3}
\end{figure}

\begin{figure}[t] 
   \centering
 \includegraphics[width = 2.5in]{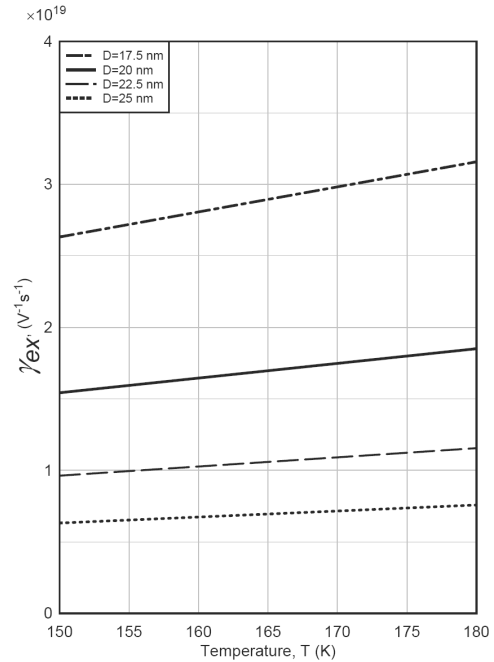} \caption{ The drag coefficient $\gamma_{ex}$ for the system of
spatially separated excitons and electrons  as a function of
temperature $T$ for the different interwell separations $D$. }
   \label{gam_lt4}
\end{figure}

Figs.~\ref{gam_lt3} and~\ref{gam_lt4}  present the results of
calculations of the drag coefficients $\gamma _{p}$ and $\gamma
_{ex}$, correspondingly, as a functions of temperature $T$ for the
different interwell separation $D$. In our calculations we used the
same interwell distances that used in the drag
experiments~\cite{Gramila,Lilly2}, namely $17.5\ \mathrm{nm}$, $20\ \mathrm{%
nm}$, $22.5\ \mathrm{nm}$ and $30\ \mathrm{nm}$. The calculations
were performed for the exciton density $n = 10^{10} \
\mathrm{cm^{-2}}$ with the parameters for the GaAs/GaAsAl quantum
wells: $m_{e} = 0.07 m_{0}$, $m_{h} = 0.15 m_{0}$, $M = 0.24 m_{0}$,
$\epsilon = 13$. All drag coefficients increase with the
temperature $T$  and decrease with the interwell separation $D$ as
$D^{-4}.$

We propose the following experiments relevant to electron-polariton
and electron-exciton drag effect. The screen with two diaphragms
covers the quantum well embedded into a semiconductor microcavity.
The laser pumping through one diaphragm generates excitons forming
the polaritons by coupling to the photons. At low temperature the
absence of the electron current will indicate the superfluidity
phase of the polaritons, while at high temperature the existence of
the electron current will indicate the drag of electrons by the
moving excitons. Therefore, the drag coefficient allows to estimate
the dragged electron current. When the electron current is induced
by the external electric field applied to the electron QW, the
photoluminescence spectrum can be measured in the other diaphragm.
At low temperature, the difference between the photoluminescence
spectrum of polaritons decay with the electric field applied to the
electrons and without the electric field will indicates that the
polaritons moved to the other place of the QW due to the drag by the
electron current. The photoluminescence spectrum in the other
diaphragm without the electric field is caused only by the diffusion
of the polaritons. Only the normal component of the polariton
subsystem will move to the other diaphragm, while the superfluid
component is not affected by the electrons. It seems like electrons
move the photons that coupled with excitons. At high temperature the
photoluminescence spectrum of the electron-hole recombination
indicates that the excitons moved to the other place of the QW due
to the drag by the electron current.

The other suggested experiment is based on the observation of the
angle  distribution of the photons escaping the optical microcavity.
At low temperature we propose to create the uniform distribution of
polaritons by the laser pumping within the microcavity. Therefore,
$\mathbf{\nabla }n_{p}=0$ and there is no polariton current. In the
absence of polariton current the average angle of the photons
escaping the optical microcavity and the perpendicular to the
microcavity is $\bar{\alpha}=0$,  because the angle distribution is
symmetrical~\cite{science}. Let us induce the electron current by
applying the electric field $\mathbf{E}$ and analyze the photon
angle distribution in the presence of the nonzero current of
polaritons along the quantum well parallel to the cavity entrained
by the electron current due to the drag effect. If $\
\mathbf{\nabla }n_{p}=0$ the polariton current is $\mathbf{i}_{p}=\gamma _{p}%
\mathbf{E}$, and according to  the definition of polariton current
we have $\mathbf{v}_{p}=\gamma _{p}\mathbf{E}/n_{p}$. Therefore, we
can obtain the average component of the polariton momentum in
the direction parallel to the Bragg mirrors of the microcavity: $\overline{%
p_{||}}=M_{eff}v_{p}=M_{eff}\gamma _{p}E/n_{p}$. Since the perpendicular
to the Brag mirrors component of the polariton momentum is given
by~$p_{\bot
}=\hbar \pi /L_{C}$ \cite{Snoke_text}, we obtain for the average tangent of the angle between the path of the
escaping photon and the perpendicular to the microcavity:
\begin{eqnarray}
\label{tan}
\overline{\tan \ \alpha }=\frac{\overline{p_{||}}}{p_{\bot }}=\frac{\gamma _{p} M_{p}L_{C}E}{\hbar \pi n_{p}}\ .
\end{eqnarray}
Note that only normal (non-superfluid) component of the polariton
subsystem will contribute to the drag coefficient $\gamma_{p}$ and
to $\overline{\tan \ \alpha }$. There will be two peaks of the
escaping photons: one at $\overline{\tan \ \alpha } \neq 0$
corresponds to the moving (dragged) normal component, and the other
one at $\overline{\tan \ \alpha } = 0$ corresponds to the superfluid
component. The analysis of the angle distributions of the photons
escaping the optical microcavity has been used in the
experiments~\cite{Amo1,Amo2}.

Let us make estimations of the parameters for the
drag effects. At the temperature $T = 4 \ \mathrm{K}$ the experiment~\cite%
{science,Balili_prb} shows that the polariton lifetime $\tau = 10 \ \mathrm{%
ps}$, and the polariton diffusion path is $l = 20 \ \mathrm{\mu m}$. The
corresponding average polariton velocity is $v_{p} = l/\tau = 2
\times 10^{6} \ \mathrm{m/s}$. Since $E= n_{p}v_{p}/\gamma_{p}$, we can estimate the electric
field $E$ corresponding to such drag effect. For
$n_{p} = 10^{10} \ \mathrm{cm^{-2}}$ and $T = 4 \ \mathrm{K}$ from Fig.~\ref{gam_lt3} for the
interwell separation $D = 17.5 \ \mathrm{nm}$ $\gamma_{p} = 2.64 \times 10^{16} \
\mathrm{V^{-1}s^{-1}}$, the corresponding electric field is $E = 3.8
\times 10^{3} \ \mathrm{V/m}$ which corresponds to the applied
voltage $V = 3.8 \times 10^{-3} \ \mathrm{V} $ at the size of the
system $d = 1 \ \mathrm{\mu m}$. Using $M_{p} = 7
\times 10^{-5} \times m_{e}$,
the length of the microcavity $L_{C} = 2 \ \mathrm{\mu m}$~\cite{science,Balili_prb}, and the estimated $\gamma_{p}$ and $E$ in~(\ref{tan}), we obtain
 for the average tangent of the angle $\alpha$:  $\overline{\tan \ \alpha} = 0.385$ and $\tan^{-1}\left(\overline{\tan \ \alpha%
} \right) = 21^{0}$.

We can conclude that at low temperatures the electron current
dragged by the polariton current is strongly suppressed.  However,
the polariton current can be dragged  by the electrons, and,
therefore, there is a transport of photons along the microcavity,
which can be observed through the angular distribution of photons
discussed above. At high temperatures, from one side, the existence
of the electric current in an electron QW indicates the exciton
current in the other QW,
 from the other side, the electron current in one QW induces the exciton current in the other QW via the drag of excitons by the electrons.
  The obtained drag coefficients allow calculate the corresponding currents in the presence of the superfluid component.
The suggested experiments allow to observe the electron-cavity
polariton drag effects and to observe directly the cavity polariton
superfluidity.

\acknowledgments

O.~L.~B., R.~Ya.~K. were supported by PSC CUNY grant 63443-0041, and
Yu.~E.~L. was partially supported by RFBR and RAS programs.


\bigskip\


\begin{thebibliography}{99}
\bibitem{text2} Here we omit weak effects of direct nonresonant interaction of
photons with 2D electron gas which will be considered elsewhere.
\bibitem{text1} If the exciton-polaritons have a finite life-time in the cavity, the
Bogoliubov dispersion is modified at small wave
vectors~\cite{Wouters}. This modification should affect the drag.
However, we consider the small  relative deviation from the
threshold pumping intensity, when the pumping
 maintains the exact balance of amplification and losses. According to Eq.~(6) in
 Ref.~[\onlinecite{Wouters}], at small  relative deviation from the
threshold pumping intensity, the spectrum of collective excitations
in  the system of exciton polaritons corresponds to the regular
 Bogoliubov approximation.
\end{thebibliography}
\end{document}